\documentclass[aps,prl,preprint,superscriptaddress]{revtex4-2}

\usepackage{amsmath}
\usepackage{graphicx}
\usepackage{color}

\begin{document}


\title{Domain Walls Stabilized by Intrinsic Phonon Modes and Engineered Defects Enable Robust Ferroelectricity in $\mathrm{HfO}_2$}

\author{Chenxi Yu}
\thanks{These authors contributed equally to this work.}
\author{Jiajia Zhang}
\thanks{These authors contributed equally to this work.}
\author{Xujin Song}
\thanks{These authors contributed equally to this work.}

\author{Dijiang Sun}
\author{Shangze Li}
\author{Fei Liu}
\email[]{feiliu@pku.edu.cn}
\author{Xiaoyan Liu}
\affiliation{School of Integrated Circuits, Peking University, Beijing, China}

\author{Wei Xi}
\email[]{xiwei@tju.edu.cn}
\affiliation{Advanced Instrumental Analysis Center, School of Chemical Engineering and Technology, Tianjin University, Tianjin, China}

\author{Jinfeng Kang}
\email[]{kangjf@pku.edu.cn}
\affiliation{School of Integrated Circuits, Peking University, Beijing, China}




\begin{abstract}
    Ferroelectric $\mathrm{HfO}_2$ has attracted extensive research interest for its applications in AI era. 
	The domain walls play a crucial role in phase structure stabilization and polarization switching of ferroelectric $\mathrm{HfO}_2$, however, a thorough understanding is still lacking. 
	Here, we developed a unified framework based on phonon mode expansion to systematically study the effects of phonon modes and defects on domain wall structures. 
	Using this approach combined with first-principle calculations, we revealed that the interface phonon modes play a key role in stability of domain walls; defects pin and stabilize ferroelectric domains, which in turn stabilizes the metastable orthorhombic phase and facilitates polarization switching. This provides an insight from the microscopic physics origin into the enhanced ferroelectricity in $\mathrm{HfO}_2$ by doping and defect engineering. Furthermore, the theoretically predicted domain structures and defect distributions were observed in La-doped $\mathrm{HfO}_2$ ferroelectric films by EELS and STEM experiments, which confirms the validity of our findings. 
\end{abstract}


\maketitle


Since the discovery in 2011 \citep{RN71}, $\mathrm{HfO}_2$-based ferroelectrics have attracted much research interests for its CMOS-compatibility \citep{RN86,RN149,RN76} (complementary metal oxide semiconductor), strong ferroelectricity in nano-scale \citep{RN149,RN77,RN74,RN87} which make it a promising candidate for applications in non-volatile memory and compute-in memory devices in artificial intelligence (AI) era. 
Tremendous efforts have been made to explore the physical origin of ferroelectricity in $\mathrm{HfO}_2$-based ferroelectrics, including the formation of ferroelectric orthorhombic phase (OIII phase) \citep{RN60}, the formation of rhombohedral phase \citep{RN2,RN215} and the ferroelectricity induced by ordered oxygen vacancies \citep{RN6,RN319}. 
Most of the existing experiments and theoretical works support the formation of OIII phase as the origin of ferroelectricity in $\mathrm{HfO}_2$ thin films \citep{RN328,RN22,RN169,RN318}. 
However, OIII phase is a meta-stable phase under ambient conditions. The ground state is archieved in monoclinic phase (M phase) from both theory \citep{RN60,RN64} and experiments \citep{RN107}. Researchers found that there are various factors existing in thin films that can stabilize the ferroelectric OIII phase, including the strain \citep{RN94,RN365,RN44,RN382}, interface \citep{RN100,RN152} and defects \citep{RN91,RN37,RN117,RN8,RN160,RN65,RN66}. 
Most existing researches on stabilizing factors focused on phase stability, switching barrier, kinetical stability and ferroelectric cycling behaviors, which are studied from the view of energy landscape. 
Although the stabilization of ferroelectric phase has been well-established from an energy landscape perspective in existing studies, its implications for device-relevant behavior, especially the dynamics of ferroelectric switching, are not well understood. 
The domain walls (DWs) are one of the key features of ferroelectrics that have significant influence on polarization switching, which is attracting more and more interests in the study of ferroelectric $\mathrm{HfO}_2$ recently \citep{RN34,RN356,RN349,RN357}. 
Therefore, the study of DWs may bridge the gap by providing a microscope view of robust ferroelectricity in $\mathrm{HfO}_2$. 
Previous researches focused on the dependence of DW stability and polarization switching path \citep{RN34,RN356,RN357}. They found that there are different types of DWs that have different stability and switching barriers. 
Recent studies found the crucial role of defects in DW stability by lowering the DW energy \citep{RN359,RN376,RN378,RN379}. 
The defects in ferroelectric $\mathrm{HfO}_2$ thin films are dopants and oxygen vacancy, the intrinsic defect that widely exist in $\mathrm{HfO}_2$ thin films \citep{RN169,RN55}. 
The defects are proved to be important impact factors in stabilizing OIII phase \citep{RN27,RN169,RN318}. 
Although there are already plenty of researches on DW stability and switching, a thorough research on DW stability and the impact of defects is still lacking. 
One of the obstacles is that the classification of DWs is a tidious task due to the 48 variants of OIII unitcells. 
These 48 OIII unitcell variants are linked by symmetry transformation under the $Fm\overline{3}m$ space group of cubic phase, therefore, the concept of pseudo-chirality is introduced to distinguish the eight OIII unitcell variants with a fixed polarization direction from six possible polarization directions \citep{RN34,RN357,RN365}. 
The large number of OIII unitcell variants is originated from the low space group symmetry of OIII phase. 
By analysing the phonon spectrum, the low-symmetry OIII unitcells can be generated from the high-symmetry cubic unitcells by the phonon modes including soft modes and hard modes \citep{RN4,RN135,RN364,RN365,RN349}. 
After analysing the phonon mode amplitudes of OIII unitcell with various pseudo-chiralities, we found that there is a one-to-one correspondence between the mode amplitudes and pseudo-chiralities, which makes it possible to represent the pseudo-chirality using phonon mode amplitudes. The mode amplitudes can be obtained by expanding the atomic coordinates of OIII unitcells using a complete basis of phonon modes. 
In this work, we developed a method based on phonon mode expansion to systematically study the effects of phonon modes and defects on domain wall structures. This method based on mode expansion is discussed in detail in our complementary work \citep{comment1}. 
Using this method, we classified all inequivalent DWs, and studied the effects of phonon modes and defects on DW structures using first-principle calculations. 
We found that the stability of unstrained DWs is determined by the polarization and pseudo-chirality of domains which are related to the phonon mode characteristics. 
The DWs act as the critical intermediary through which defects impact the stability of OIII phase by pinning the DWs and increasing DW stability. 
The defects concentrated on the DW can facilitate DW switching through DW motion from the DW interface by lowering the switching barrier \citep{RN30,RN48,RN43}. 
Furthermore, we conducted EELS analysis and STEM characterization to observe the distribution of defects and the DW structures in La-doped ferroelectric $\mathrm{HfO}_2$ thin films to confirm our findings. 

Firstly, we defined eight pseudo-chirality numbers of OIII unitcell to represent the pseudo-chirality of domains in DWs. 
The pseudo-chirality has a one-to-one correspondence to the mode amplitudes, therefore, we made the mode expansion of atom displacements in OIII phase relative to that of cubic phase. 
We calculated the phonon spectrum of cubic phase using the conventional cubic unitcell, and extracted all the modes on the $\Gamma$ point of cubic conventional cell. 
We found that the modes of the sixth branch can uniquely determine the pseudo-chirality of OIII unitcell, which is shown in Equation \ref{eq:expansion-OIII-variants-Q6}. 
The sixth branch has degeneracy of six and irreducible representation of $X_5$, whose mode amplitudes are denoted by $Q_{6zx}, Q_{6xy}, Q_{6yz}, Q_{6yx}, Q_{6zy}, Q_{6xz}$, where the first subscripts $x, y, z$ indicate that the phonon mode is folded from X, Y and Z point in the primitive Brillouin zone and the second subscripts $x, y, z$ indicate the direction of atom displacements. 

\begin{equation}
	\begin{aligned}
		Q_6(\text{OIII,primary}) = 0.052 \left(\epsilon_1 Q_{6zx} + \epsilon_2 Q_{6xy} + \epsilon_3 Q_{6yz}\right) \\
		Q_6(\text{OIII,conjugate}) = 0.052 \left(\epsilon_1 Q_{6yx} + \epsilon_2 Q_{6zy} + \epsilon_3 Q_{6xz}\right)
	\end{aligned}
	\label{eq:expansion-OIII-variants-Q6}
\end{equation}

These $Q_6$ modes appear in two triplets which are $C_4$-conjugated and denoted the primary and conjugate triplets. 
Using the two types of triplets and three signs $\epsilon_1, \epsilon_2, \epsilon_3$, we can define eight pseudo-chirality numbers in Table \ref{tab:OIII-chirality-number} which are related to phonon modes of OIII phase by the mode expansion expression of Equation \ref{eq:expansion-OIII-variants-Q6}. 
The detailed discussion on the mode expansion approach is shown in our complementary work \citep{comment1}. 

\begin{table}[htbp]
    \centering
	\caption{Signs and conjugate pairs in mode expansion of OIII and the definition of (pseudo-) chirality number \label{tab:OIII-chirality-number}}{%
    \begin{tabular}{ccccc}
		Chirality Number & Primary/Conjugate & $\epsilon_1$ & $\epsilon_2$ & $\epsilon_3$ \\
        \hline
		0 & P & + & + & + \\
		1 & P & + & - & - \\
		2 & P & - & + & - \\
		3 & P & - & - & - \\
		4 & C & + & + & + \\
		5 & C & + & - & - \\
		6 & C & - & + & - \\
		7 & C & - & - & + \\
    \end{tabular}}{}
\end{table}

\begin{figure}
    \includegraphics[scale=0.7]{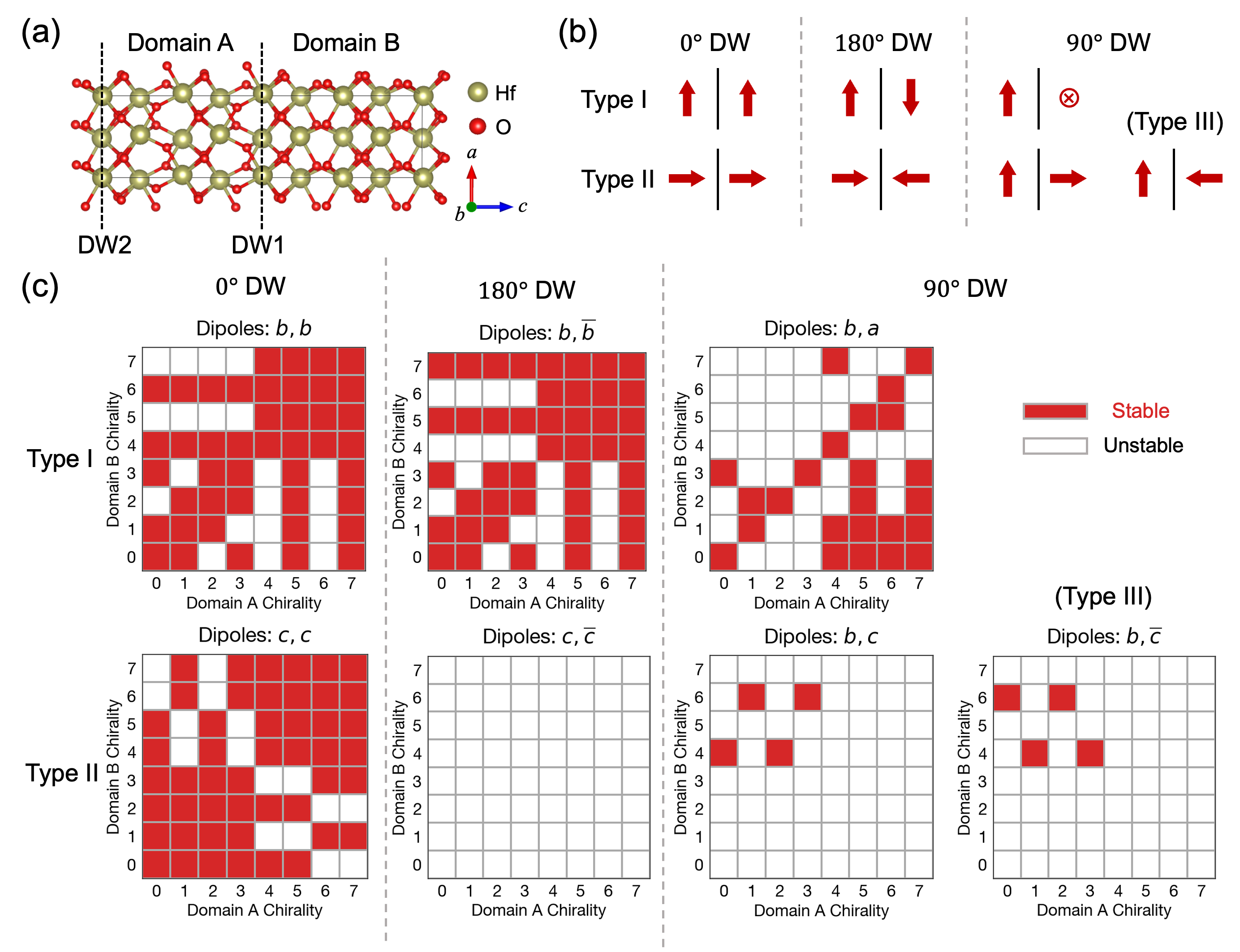}
    \caption{Dependence of OIII domain wall stability on interface phonon modes. 
	(a) The domain wall model with two domains. 
	(b) Polarization configuration, and different types of $0^\circ, 180^\circ, 90^\circ$ domain walls. 
	(c) Dependence of OIII domain wall stability on interface phonon modes, or pseudo-chirality number at interface. 
	The x-axis and y-axis of each stability map are the pseudo-chirality number of domain A and B in subfigure a. 
	The figure title of each stability map shows the dipole configuration of domain A and B. 
	The overlined axis is the compact notation for the reversed axis direction, \textit{e.g.} $\overline{c}$ means $-c$. 
    \label{fig:domain-wall-stability}}
\end{figure}

Using the pseudo-chirality number and polarization direction, we enumerated all the inequivalent DWs and studied effect of modes on their stability. We constructed the DW model comprised of two domains as shown in Fig. \ref{fig:domain-wall-stability}a. The DWs with interface along one of (100), (010) and (001) directions have seven inequivalent polarization configurations as shown in Fig. \ref{fig:domain-wall-stability}b, with two types of $0^\circ$ DWs, two $180^\circ$ DWs and three $90^\circ$ DWs. 
We used first-principle calculations by VASP software \citep{RN109,RN110} to do structure relaxation of these DWs. 
Some DWs are not stable after structure relaxation, which can be classified into these cases: the transition into single-domain structure; the transition from orthorhombic phase to monoclinic phase or tetragonal phase; the motion of DW by half-cell, increasing or decreasing domain size by half-integer; and the destruction of the whole superstructure. 
We found that the stability of DWs is related to the dipole configuration and the pseudo-chirality number of two domains, which are directly related to interface phonon modes. 
Fig. \ref{fig:domain-wall-stability}c shows the dependence of DW stability on dipole configuration and pseudo-chirality number of two domains. 
From these stability maps, we can summarize some stability rules: 

\begin{enumerate}
	\item The number of stable DWs depends on the angle between polarization vectors, following the order: $0^\circ > 180^\circ > 90^\circ$. 
	\item The $180^\circ$ DW with two polarization vectors parallel to normal vector is not stable. 
	\item The $90^\circ$ DW with two $Q_1$ modes parallel to each other is not stable. 
	\item Other individual unstable cases. 
\end{enumerate}

The first rule has these origins: the $0^\circ$ DW has less dipole interaction energy than $180^\circ$ DW and $90^\circ$ DW; $90^\circ$ DW has larger lattice mismatch than $0^\circ$ DW and $180^\circ$ DW; and different mode interactions in $180^\circ$ DW and $90^\circ$ DW. 
The second rule is related to the DW charges. The $180^\circ$ DW with two polarization vectors parallel to normal vector has the largest interface charge density, therefore, it is the most unstable DW. 
The $90^\circ$ DW with one polarization vector normal to interface has half the interface charge density as the most unstable $180^\circ$ DW, as a consequence, it is the second most unstable DW. 
However, we observed many $90^\circ$ DWs in experiments \citep{RN257}, including that with one polarization vector normal to interface, which seems contradictory to our finding. 
In fact, these DWs are stabilized by defects, which modulate the mode interactions at interface and reduce interface charges \citep{RN359}. 
The third rule is related to interface mode interactions, causing the instability at these regions where the two $Q_1$ modes are parallel to each other: the left-upper corner in the stability map of dipole configuration $b,a$; and the left-upper, left-bottom and right-bottom corners of dipole configuration $b,c$ and $b,\overline{c}$. 
This rule is not strictly obeyed by DWs with defects which modulate the mode interactions, but the DWs following this rule can be stabilized by more defect arrangements than those not following this rule. 

\begin{figure}
    \includegraphics[scale=0.7]{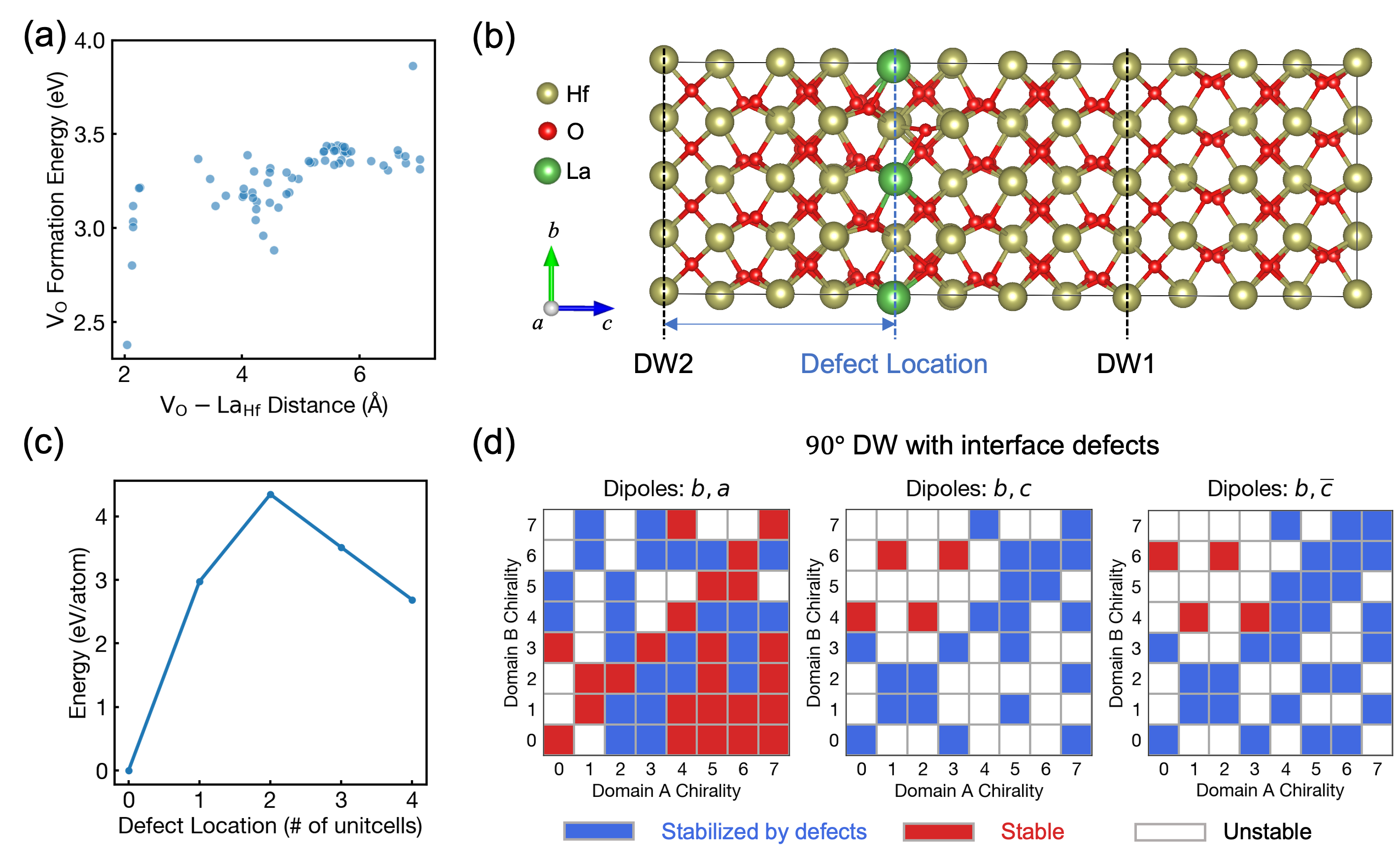}
    \caption{Effects of defects on DWs. 
	(a) Dependence of the formation energy of La substitutional defect $\mathrm{La}_\mathrm{Hf}$ and oxygen vacancy $\mathrm{V}_\mathrm{O}$ pairs on their distance, exhibiting a positive correlation, with the minimum formation energy achieved at the minimum distance. 
	(b) The structure model used to study the dependence of DW energy on the distribution of defects. We put $\mathrm{La}_\mathrm{Hf}$ and $\mathrm{V}_\mathrm{O}$ defects in one of the domains with varying distances to the DW interface. 
	(c) Dependence of energy of the model in subfigure b on the distance from the defects to DW interface. The energy is the lowest when the defects are at the interface, which indicates that the DW interfaces are more likely to form in the area with high defect density. 
	(d) Dependence of stability of $90^\circ$ DW with defects on interface phonon modes, or pseudo-chirality number at interface. We use different colors to distinguish the DWs that are stable with or without defects (blue) and the DWs that are unstable without defects but stable with defects (green). 
    \label{fig:domain-wall-defect}}
\end{figure}

From the above analysis of phonon mode effect, we found that the defects play an important role on the stabilization of DW besides the modes. 
Due to the trivalent nature of La, the doping of La introduces oxygen vacancies $\mathrm{V}_\mathrm{O}$ into $\mathrm{HfO}_2$ for charge compensation. 
We calculated the formation energy of $\mathrm{La}_\mathrm{Hf}$ and $\mathrm{V}_\mathrm{O}$ defect pairs versus the distance between the defect pairs as shown in Fig. \ref{fig:domain-wall-defect}a, and found that the formation energy has a positive correlation with the defect distance, which indicates that the $\mathrm{V}_\mathrm{O}$ has a high probability to form near the La dopant. 
Next, we studied the distribution of defects inside the domain structure. We constructed the DW model comprised of two domains as in Fig. \ref{fig:domain-wall-defect}b, with one of the domains have $\mathrm{La}_\mathrm{Hf}$ and $\mathrm{V}_\mathrm{O}$ defects. The DW energy versus the distance from defects to DW interface in Fig. \ref{fig:domain-wall-defect}c shows that the lowest energy configuration is archieved when the defects are at the interfaces, or the defects have a pinning effect on the DW interfaces. 
Based on this defect pinning effect, we calculated the stability map of DWs with defects by adding defects at the DW interfaces. We chose the $90^\circ$ DWs which are the most unstable without defects to calculate, and the results are shown in Fig. \ref{fig:domain-wall-defect}d that the defects can stabilize most of the DWs that are not stable without defects. 
It is worth to mention that the stabilized $90^\circ$ DW by certain arrangement of defects may not be stable by other arrangements of defects. 
This fact may be caused by the artifacts of three-dimensional periodic condition used in calculation, even though we made $2 \times 2$ supercell in two interface directions, the structure is still not large enough for simulation. 
Also, this fact implies that the ferroelectricity of $\mathrm{HfO}_2$ is intertwined to defects \citep{RN32,RN6,RN18,RN8,RN162,RN251}. 

\begin{figure}
    \includegraphics[scale=0.7]{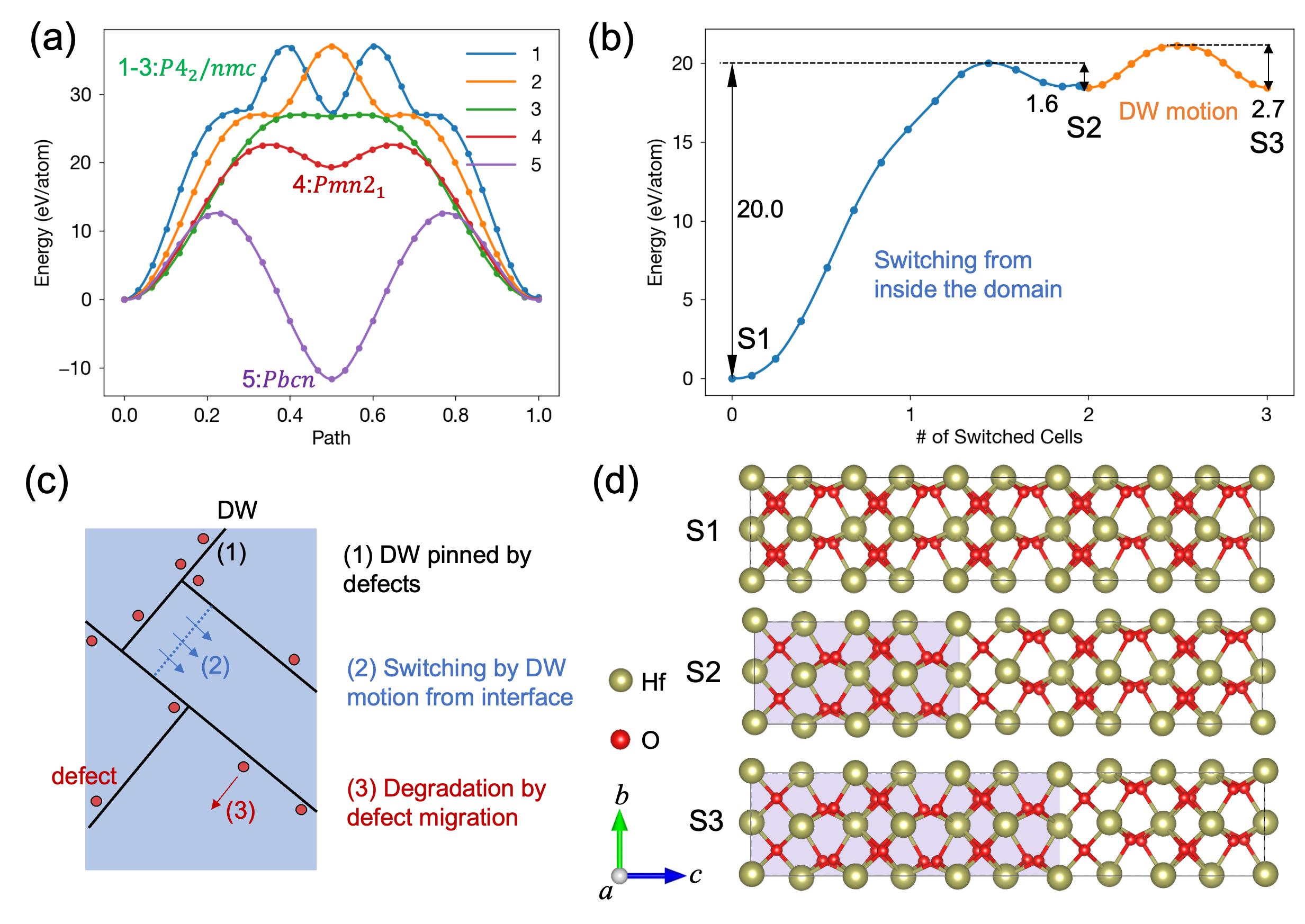}
    \caption{Ferroelectric switching mechanism based on DW motion. 
	(a) Summary of five switching paths of OIII phase. 
	(b) The switching path from S1 to S2, and the path from S2 to S3. The two paths correspond to the switching from inside the domain and the DW motion. The structures of S1 to S3 are shown in subfigure c. 
	(c) Schematic of proposed switching mechanism. 
	(d) Structures in the switching paths in subfigure b: S1, the single-domain ground state; S2, a two-cell-reversed state; S3, a three-cell-reversed state. 
    \label{fig:domain-wall-switching}}
\end{figure}

Subsequently, we studied the switching of DW using first-principle calculation. 
Using mode expansion method, we enumerated all the inequivalent switching paths and summarized switching mechanisms of OIII phase in Fig. \ref{fig:domain-wall-switching}a \citep{comment1}. Most of the switching paths follow the tetragonal ($P4_2/nmc$) mechanism of paths numbered one to three in the figure \citep{comment1}, therefore, we used this mechanism to study the switching of DW. 
We constructed single-domain structure and structures with some of the cells reversed shown in Fig. \ref{fig:domain-wall-switching}d, and calculated their transition paths in Fig. \ref{fig:domain-wall-switching}b. 
The first path is from the ground state of single-domain structure to the structure with two cells switched, for the structure with one unitcell switched is not stable. This path can be seen as the ferroelectric switching starting from inside the domain, due to the periodic boundary condition used in calculation. The second path is from the single-domain structure with two cells switched to that with three cells switched, corresponding to the DW motion. We can see that the barrier is very large to switch from single domain, but becomes lower than the switching barrier between OIII phases once two cells are switched. These facts have three origins: the cells switched starting from single domain should be at least two, which raises the barrier, for the structure with one cell switched is not stable; the switching from single domain introduces strain; the DW motion involves the interaction of modes at the interface only, compared with the switching path of OIII phase which requires the collective motion of all the modes in cells. We conjecture that the switching of DW mostly starts from the interface of DWs, because the defects distributed near the DW interface can lower the switching barrier significantly, similar to the case of OIII switching \citep{RN30,RN48,RN43}. 

\begin{figure}
    \includegraphics[scale=0.7]{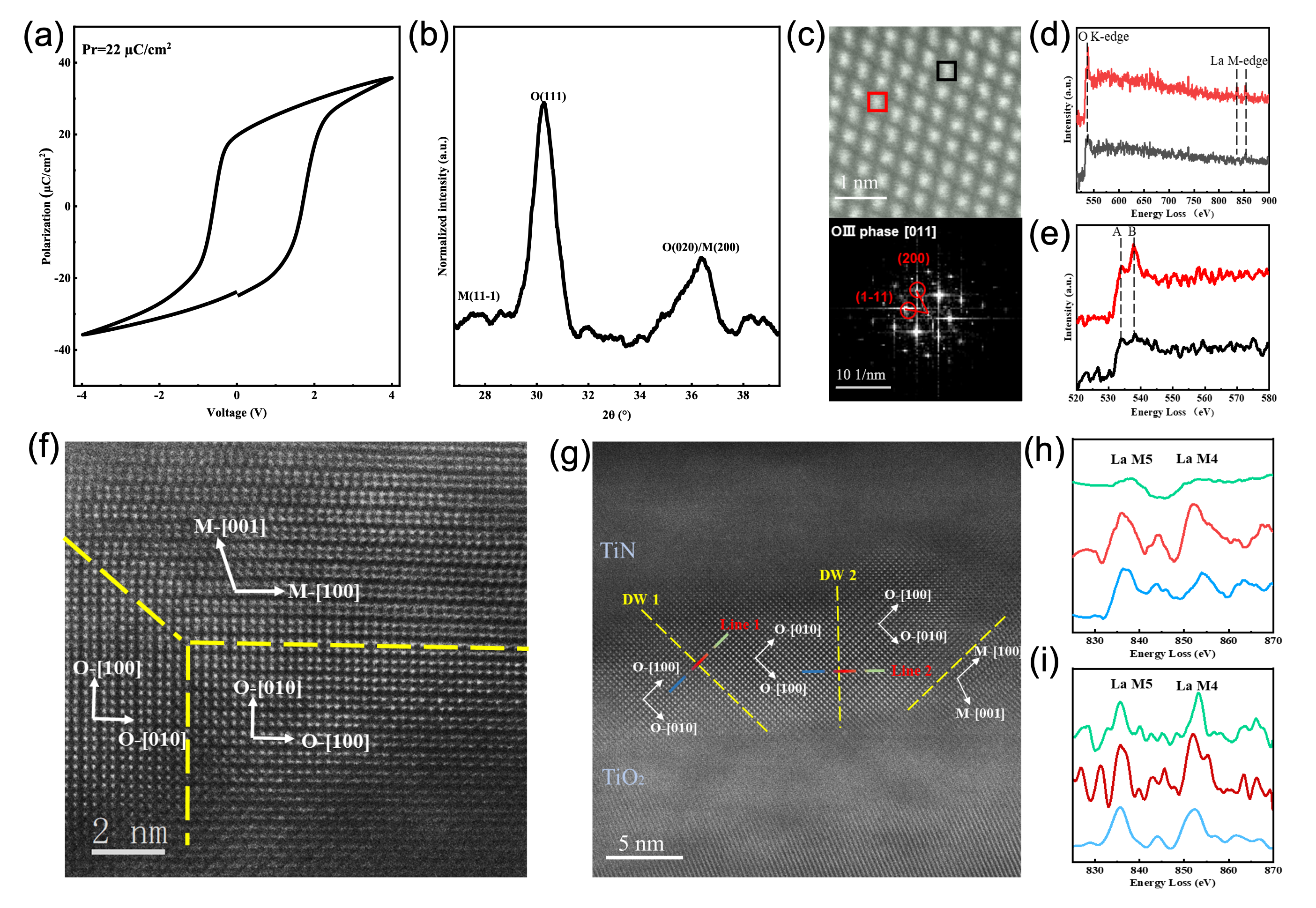}
    \caption{Structural characterization and defect analysis of HLO films. 
	(a) P-V hysteresis loop of the $\mathrm{TiN}/\mathrm{HLO}/\mathrm{TiO}_2$ stack showing a low coercive voltage of $2V_c=2.3\mathrm{V}$ and the remanent polarization of $P_r = 22 \mu \mathrm{C}/\mathrm{cm}^2$. 
	(b) GIXRD patterns of HLO film. 
	(c) STEM-HADDF image and corresponding fast-Fourier-transformation (FFT) patterns. 
	(d) EELS spectra and 
	(e) O-K edges obtained from the region marked with red line and black line in (c). 
	(f) Atomic structures of HLO film. 
	(g) The coexistence of O domains with DWs in a single grain. La-M edges of 
	(h) Line1 and 
	(i) Line2 obtained from the DW1 and DW2. Each spectrum color corresponds to similiar color position marked in (g).
    \label{fig:exp}}
\end{figure}

To visualize the domain structure and verify the stabilizing effect of defects such as oxygen vacancies, lanthanum-doped hafnium oxide (HLO) films were fabricated on $\mathrm{TiO}_2$ films by atomic layer deposition (ALD). Fig. \ref{fig:exp}a shows the typical polarization–voltage (P–V) loops of the $\mathrm{TiN}/\mathrm{HLO}/\mathrm{TiO}_2$ structure measured at 10 kHz, which exhibits the excellent ferroelectric properties of HLO. Grazing incidence X-ray diffraction (GIXRD) was employed to determine the phase composition of HLO films. 
The diffraction pattern (Fig. \ref{fig:exp}b) demonstrates a dominant (111)-oriented OIII phase, which is crucial for ferroelectricity, alongside a minor M phase. 
To elucidate the elemental distribution of OIII phase, scanning transmission electron microscopy (STEM) high-angle annular dark field (HAADF) imaging combined with electron energy loss spectroscopy (EELS) analysis were performed along the O-[011] zone axis. 
Quantitative analysis of La and O energy loss spectra (Fig. \ref{fig:exp}d) reveals distinct compositional variations: black regions correspond to uniform La doping, while red regions indicate La-enriched domains. 
The corresponding background-subtracted O-K edge (Fig. \ref{fig:exp}e) further represent the fine structural differences driven by $\mathrm{V_O}$. 
Oxygen vacancies modify the local crystal symmetry and electronic structure by influencing the hybridization between O-2p and Hf-d/La-d orbitals. 
The presence of $\mathrm{V}_\mathrm{O}$ introduces shallow donor states below the conduction band, altering the relative intensities of peak A ($\sim 534 \mathrm{eV}$) and peak B ($\sim 537 \mathrm{eV}$) \citep{RN381}. 
The significantly lower intensities of A/B peaks in the La-enriched (red) region suggest a correlation between higher La doping and increased oxygen vacancy concentration. This relationship can be understood through the charge compensation mechanism: the substitution of $\mathrm{Hf}^{4+}$ by $\mathrm{La}^{3+}$ creates a net negative charge that must be compensated by oxygen vacancies to maintain charge neutrality.

Atomic-resolution STEM imaging along the O-[001] zone axis was conducted to further examine the phase distribution and domain structure. 
This orientation is selected to enhance the visibility of oxygen displacements, facilitating the identification of polymorphs and domain boundaries. 
As shown in Fig. \ref{fig:exp}f-g, OIII ferroelectric domains are separated by $90^\circ$ DW, with a minor fraction of M phase domains coexists at the interfaces. 
Fig. \ref{fig:exp}g reveals three distinct OIII ferroelectric domains separated by two types of domain walls. 
The DW1 is oriented along the O-[100]/O-[010] plane, while DW2 lies in the O-[110] plane. 
The stability of these domain walls is enhanced by the preferential segregation of La dopants and associated $\mathrm{V}_\mathrm{O}$. 
The EELS line scans of La distribution (Fig. \ref{fig:exp}h-i) reveal that both DW1 and DW2 exhibit significantly higher La concentrations compared to the bulk domains. 
According to the lowest-energy configuration, the formation of $\mathrm{La}_\mathrm{Hf}$-$\mathrm{V}_\mathrm{O}$ complexes enhances the stability of the DW interfaces. 
These initial defects are randomly distributed. 
This stands in contrast to the previous research results in fatigued structures, where the formation of periodic $\mathrm{V}_\mathrm{O}$ led to degraded ferroelectricity[57]. 
This comparison indicates that a randomized vacancy distribution, rather than a periodic one, is crucial for stabilizing domain walls and maintaining robust ferroelectric switching. 

In summary, we systematically studied the DW stability and the effects of defects on DWs in ferroelectric La-doped $\mathrm{HfO}_2$. 
We developed a method based on phonon mode expansion to classify the symmetrically inequivalent DWs, and carried out first principle calculations. We found that the oxygen vacancies are distributed near the La dopants and the defects have a pinning effect on the DW interface (Fig. \ref{fig:domain-wall-switching}c), which is confirmed by EELS and STEM experiments (Fig. \ref{fig:exp}). 
The stability of DWs depends on pseudo-chiralities of domains, or the interface phonon modes. 
The defects can increase the overall stability of DWs with varying interface phonon modes. 
The DW structures stabilized by defects lead to the robust ferroelectricity in La-doped $\mathrm{HfO}_2$, and also facilitate ferroelectric switching through domain wall motion mechanism shown in Fig. \ref{fig:domain-wall-switching}c. 
The switching barrier can be lowered by the defects concentrated near the DW interface. 
These findings provide the microscopic origin for the enhanced ferroelectricity in $\mathrm{HfO}_2$ by doping and defect engineering, which may provide guidance for the device optimization, especially the ferroelectric cycling behaviors which may be explained by the interplay between defects and DWs. There are experimental evidences showing that the defects can move during field cycling \citep{RN32,RN162,RN65}. We conjectured that the defects may have an ordered distribution after a large number of field cycling, accompanied by the generation of more oxygen vacancies. This redistributed defects may lose the pining and stabilization effects, leading to the transition from OIII phase to a monoclinic phase with defects which is polar but not switchable \citep{RN319,RN1}. This defect-induced phase transition may be one of the causes of the imprint effect and fatigue effect. 

\pagebreak

\appendix*

\section{First Principle Method}

Structure relaxation of DW model in Fig. \ref{fig:domain-wall-stability}a was carried out by first-principle calculation implemented in VASP software \citep{RN109,RN110}. 
Chemical species were treated using the projector-augmented plane wave method \citep{RN373}, with $5p^6 6s^2 5d^2$ valence electron configuration for Hf and $2s^2 2p^4$ for O. 
The plane wave cutoff energy was 600 eV and the Brillouin zone was sampled with gamma-centered k-point meshes with resolution 0.04. 
Exchange-correlation functional was approximated by the Perdew-Burke-Ernzerhof functional (PBE) \citep{RN148} in generalized gradiant approximation (GGA). 
The atom force convergence threshold was set to $0.01 \mathrm{eV}/\text{Å}$ and energy convergence threshold was set to $10^{-6} \mathrm{eV}$. 

DW model has one unitcell in a and b axis directions, and four unitcells in c axis direction, with two unitcells belonging to domain A and the other two belonging to domain B. 
The DW model with defects has two unitcells in a and b axis directions, and four unitcells in c axis direction, with two unitcells belonging to domain A and the other two belonging to domain B. 
The DW model has two domain wall interfaces as shown in Fig. \ref{fig:domain-wall-stability}a. 
25\% of the Hf atoms at each DW interface are substituted by La atoms, and one oxygen vacancy per two La substitutional defects is introduced to compensate charges. 
The oxygen vacancy is adjacent to one of the La dopants. 

All the structures in figures were visualized using VESTA software \citep{RN189}. 

\section{Film Fabrication}

The $\mathrm{TiN}/\mathrm{HLO}/\mathrm{TiO}_2$ heterostructure investigated in this work was previously reported by our group \citep{RN361} (Fig. S1). 
Initially, an 8-nm-thick $\mathrm{TiO}_2$ was deposited via ALD using $\mathrm{TiCl}_4$ precursor and $\mathrm{O}_3$ as oxidant at $300^\circ\mathrm{C}$. 
The $\mathrm{TiO}_2$ film was crystallized through rapid thermal annealing (RTA) at $600^\circ\mathrm{C}$ for 30 s in $\mathrm{N}_2$ atmosphere. 
Following this process, an 8-nm-thick La-doped $\mathrm{HfO}_2$ film was deposited using atomic layer deposition (ALD) with $\mathrm{Hf}[\mathrm{N}(\mathrm{C}_2\mathrm{H}_5)\mathrm{CH}_3]_4$, $\mathrm{La}(\mathrm{C}_{11}\mathrm{H}_{19}\mathrm{O}_2)_3$ precursors and $\mathrm{O}_3$ as oxidant at $300^\circ\mathrm{C}$. 
The Hf:La ratio was controlled to be 11:1 by adjusting the ALD cycle ratio of $\mathrm{HfO}_2$ and $\mathrm{La}_2\mathrm{O}_3$, as confirmed by energy dispersive spectroscopy (EDS) analysis (Fig. S2). 
Subsequently, a 40-nm-thick $\mathrm{TiN}$ film was deposited by physical vapor deposition (PVD). 
Finally, the $\mathrm{TiN}/\mathrm{HLO}/\mathrm{TiO}_2$ stack underwent RTA at $600^\circ\mathrm{C}$ for 30 s in $\mathrm{N}_2$ atmosphere to promote crystallization. 

\section{Experimental Characterization}

P-V hysteresis loop of the $\mathrm{TiN}/\mathrm{HLO}/\mathrm{TiO}_2$ stack was measured using 10 kHz triangular pulse with amplitude of 4 V applied to the TiN top electrode while the $\mathrm{TiO}_2$ bottom layer was grounded.

Structural characterization of the stack was performed using grazing-incidence-angle XRD (GIXRD) instrument (Smartlab) operated at 150 mA and 50 kV. The incidence angle was fixed at $2^\circ$, and scanning speed was $4^\circ/\mathrm{min}$. The overall and atomic structures of the HLO layers were examined by high-angle annular dark-field scanning transmission electron microscopy (HAADF-STEM) using a Cs-corrected microscope operated at 300 kV (JEM-ARM300F2 FHP, JEOL). The collection angle range of the HAADF and annular bright-field (ABF) detectors were 64-180 mrad and 12-24 mrad, respectively. Atomic-scale HAADF-STEM images were processed using a Wiener filtering method (Radial Wiener Filter) to reduce random background noise. Electron energy-loss spectroscopy (EELS) spectral imaging data for the HLO samples were recorded over an energy range of 500–900 eV, which encompassed the O K and La M edges, using a Gatan EELS spectrometer. The apparatus had an entrance aperture of 5 mm and an energy dispersion of 0.3 eV per channel. 

\begin{acknowledgments}
	This work is supported by the National Natural Science Foundation of China (Grants No. T2293703, No. T2293700).
\end{acknowledgments}

\bibliography{ref}

\clearpage
\widetext
\begin{center}
\textbf{\large Supplemental Materials: Domain Walls Stabilized by Intrinsic Phonon Modes and Engineered Defects Enable Robust Ferroelectricity in $\mathrm{HfO}_2$}
\end{center}

\setcounter{section}{0}
\setcounter{page}{1}
\setcounter{secnumdepth}{2}
\appendix
\makeatletter

\renewcommand{\theequation}{S\arabic{equation}}
\renewcommand{\thefigure}{S\arabic{figure}}
\setcounter{figure}{0}

The Supplementary material includes STEM and EDS experiment results of our film. 

\begin{figure}
    \includegraphics[scale=1.0]{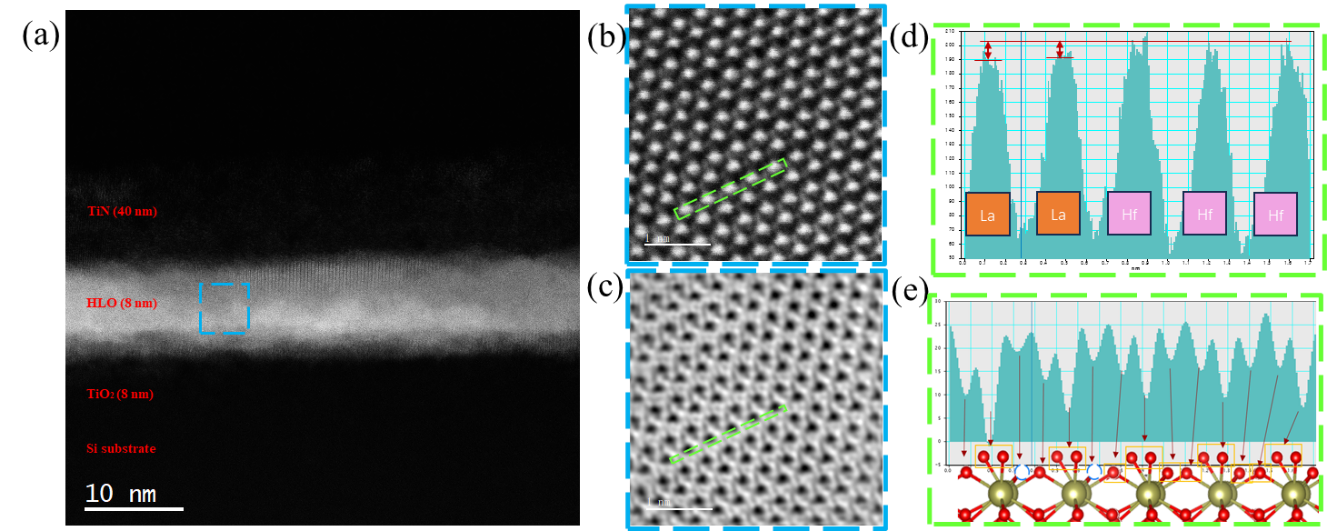}
    \caption{Structural characterization of HLO film. 
	(a) Cross-sectional HADDF image of $\mathrm{TiN}/\mathrm{HLO}/\mathrm{TiO2}$ stack. 
	The HADDF (b) and ABF (c) images projected along O-[011] zone axis. The green lines indicate the extraction contrast positions. 
	(d) The contrast curve is used to identify the positions of La atoms in the HLO film. 
	(e) The contrast curve is used to identify the positions of oxygen atoms in the HLO film. 
    \label{fig:supp-1}}
\end{figure}

\begin{figure}
    \includegraphics[scale=0.5]{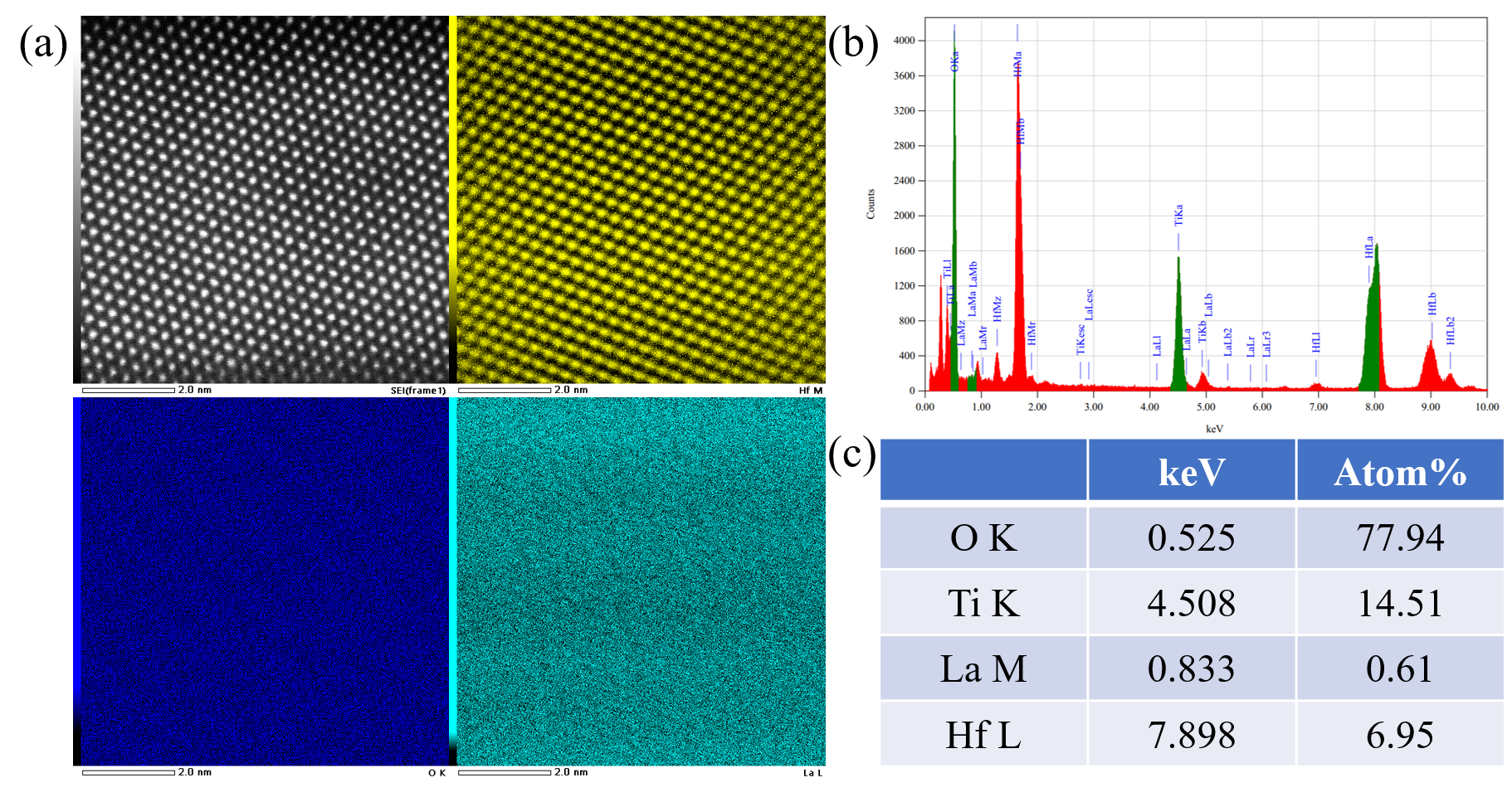}
    \caption{Element distribution of HLO film. 
	(a) EDS elemental mapping analysis. 
	(b) X-ray spectra with O, Ti, La and Hf elements corresponding line. 
	(c) The proportion of element contents in (b).
    \label{fig:supp-2}}
\end{figure}

Characterization of the $\mathrm{TiN}/\mathrm{HLO}/\mathrm{TiO}_2$ structure and element distribution was conducted using STEM (Fig. \ref{fig:supp-1}) and EDS (Fig. \ref{fig:supp-2}). 
The STEM-HADDF images indicated that the $\mathrm{TiN}$-$\mathrm{HLO}$-$\mathrm{TiO}_2$ heterointerface was coherent. 
To further investigate the element distribution relationships within HLO films, atomic-resolution STEM-HADDF and STEM-ABF images were conducted along the O-[011] zone axis (Fig. \ref{fig:supp-1}b-c). 
STEM-HADDF technology is highly sensitive to the atomic number (Z) of the analyzed material. 
Elements with higher Z values exhibit stronger electron beam interactions with atomic nuclei, resulting in greater electron scattering at high angles. Due to the contrast difference between La (Z=57) and Hf (Z=74), La-enriched locations are thus identified (Fig. \ref{fig:supp-1}d). 
The STEM-ABF image is formed by weakly scattered electrons and has the advantage of higher sensitivity to light elements such as oxygen. It can also display heavier atoms. 
The ABF image corresponding to the position of the oxygen atom in the HADDF image is used to determine the distribution of oxygen vacancies (Fig. \ref{fig:supp-1}c). 
As shown in Fig. S1e, the number of oxygen atoms at the lanthanum-enriched position decreases, which also corresponds to the position with the lowest formation energy of oxygen vacancies in the calculated results. 

\begin{figure}
    \includegraphics[scale=1.0]{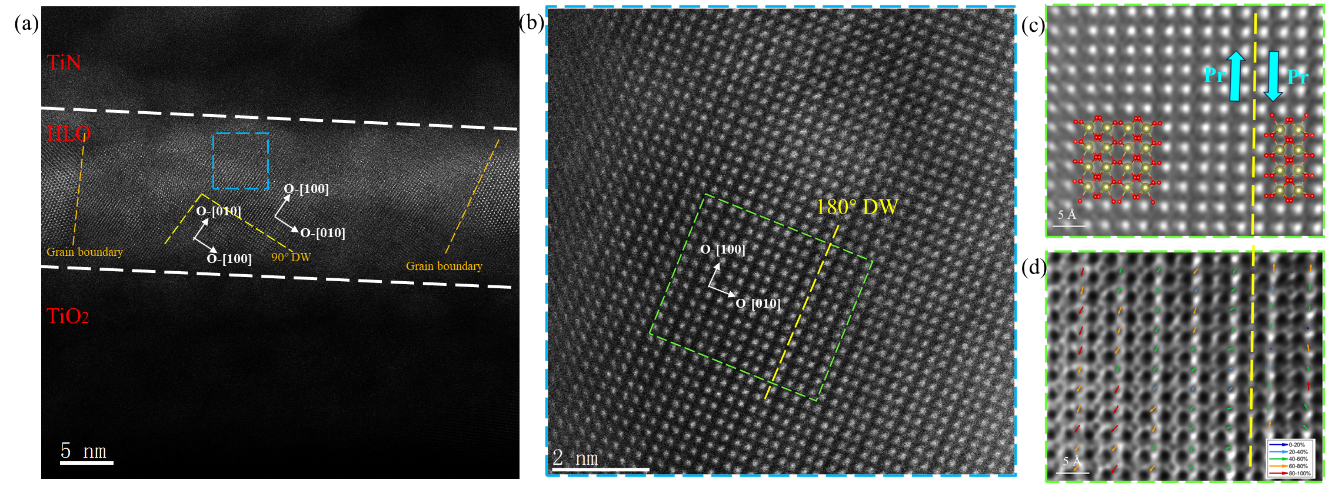}
    \caption{STEM-images of the HLO film projected along O-[001] zone axis. 
	(a) HLO grains can consist of different portions of ferroelectric domains and DWs. 
	(b) HADDF image acquired from the blue square area in (a). 
	(c) Enlarged drawings obtained from the selected green areas highlighted in (a), where the arrows indicate the corresponding direction of polarization. 
	(d) Three-fold coordinated oxygen atom displacement vectors and $180^\circ$ DW. 
    \label{fig:supp-3}}
\end{figure}

Furthermore, STEM observations of HLO crystal grain reveal different domain structures partitioned by $90^\circ$ and $180^\circ$ DWs (Fig. \ref{fig:supp-3}a). 
The oxygen atomic columns are clearly resolved in the STEM-ABF image (Fig. \ref{fig:supp-3}d). 
Specially, the three-fold coordinated atomic columns on the left shift along the O-[100] direction, while the opposing columns on the right shift along the O-$[\overline{1}00]$direction. 
This anti-parallel displacement confirms the existence of a $180^\circ$ DW, as indicated by the dashed white line in Fig. \ref{fig:supp-3}c. 

\end{document}